\documentclass[twocolumn,nofootinbib,showpacs,epsfig,prl,reprint]{revtex4}

\usepackage{amsmath}
\usepackage{amsfonts}
\usepackage{amssymb}
\usepackage{graphicx}
\usepackage{color}

\usepackage{hyperref}
\hypersetup{colorlinks,linkcolor=blue,filecolor=green,urlcolor=blue,citecolor=blue}

\newcommand{\be}{\begin{equation}}
\newcommand{\ee}{\end{equation}}

\newcommand{\sysb}{\left\{\begin{array}}
\newcommand{\syse}{\end{array}\right.}

\newcommand{\ag}[1]{}

\begin{document}

\title{Thermal and non-thermal signatures of the Unruh effect in Casimir-Polder forces}

\author{Jamir Marino,${}^{1, 2, 3}$ Antonio Noto,${}^{4, 5}$ and Roberto Passante${}^{4}$}

\affiliation{$^1$ SISSA --- International School for Advanced Studies, via Bonomea 265, 34136 Trieste, Italy \\
$^2$ INFN --- Istituto Nazionale di Fisica Nucleare, sezione di Trieste \\
$^3$ Institute for Theoretical Physics, University of Innsbruck, Technikerstrasse 21a, A-6020 Innsbruck, Austria \\
$^4$ Dipartimento di Fisica e Chimica, Universit\`{a} degli Studi di Palermo and CNISM, Via Archirafi 36, I-90123 Palermo, Italy \\
$^5$ Universit\'e Montpellier  2, Laboratoire Charles Coulomb UMR 5221- F-34095, Montpellier, France}

\begin{abstract}
We show that Casimir-Polder forces between two relativistic uniformly accelerated atoms exhibit a transition from the short distance thermal-like behavior predicted by the Unruh effect, to a long distance non-thermal behavior, associated with the breakdown of a local inertial description of the system. This phenomenology extends the Unruh thermal response detected by a single accelerated observer to an accelerated spatially extended system of two particles, and we identify the characteristic length scale for this crossover with the inverse of the proper acceleration of the two atoms. Our results are derived separating at fourth order in perturbation theory the contributions of vacuum fluctuations and radiation reaction field to the Casimir-Polder interaction between two atoms moving in two generic stationary trajectories separated by a constant distance, and linearly coupled to a scalar field. The field can be assumed in its vacuum state or at finite temperature, resulting in a general method for the computation of Casimir-Polder forces in stationary regimes.
\end{abstract}


\date{\today}

\maketitle

\emph{Introduction.} Recent significant experimental progresses in measuring fluctuation induced forces with unprecedented level of accuracy \cite{Mohideen2009,BookReview}, from the microscopic to the macroscopic level, have strongly renewed the interest in Casimir and Casimir-Polder interactions \cite{Casimir,CasimirPolder,Lamoreaux05,BW07}, even at finite temperature \cite{Barton01,OWAPSC07} and in out of equilibrium configurations \cite{APSS08}. The interest on Casimir physics spreads ubiquitously, from quantum electrodynamics towards cosmology, statistical mechanics and colloidal physics, as well as material science and nanophysics \cite{BookReview} \cite{BordagBook}.

A peculiar aspect of the quantum vacuum is that its particle content is observer dependent. The Unruh effect is a striking manifestation of this fact:
a uniformly accelerated observer in flat spacetime associates a thermal bath to the power spectrum of quantum vacuum fluctuations, with a temperature proportional to its proper acceleration \cite{Unruh76,CHM08,Fulling73,Davies75}.
Despite its conceptual importance for connections with Hawking radiation \cite{Hawking} and for its impact on  cosmology, black hole physics, particle physics, and relativistic quantum information \cite{QI}, experimental detection of the Unruh effect remains elusive, since an acceleration of the order of $10^{20}$m$/$s$^2$ would be required in order to measure a temperature of $1$K.
On the other hand, it has been recently shown that the van der Waals interaction between two accelerated atoms could allow to detect the Unruh effect with reasonable values of the acceleration \cite{NP13}.
Even if it is a well established fact that an accelerated observer perceives the vacuum field as a thermal state \cite{CHM08}, the tantalizing possibility to find non-thermal features associated to relativistic uniform acceleration would constitute a sharp representative signature of accelerated motion, beyond the Unruh thermal analogy \cite{Passante98,RS09,ZY10,MP10}.

In this Letter, we aim at bridging the fields of Casimir forces and of the Unruh effect, showing that both thermal and non-thermal features associated to a relativistic uniformly accelerated motion can be probed through the Casimir-Polder force between two accelerating atoms. In order to inspect the hallmarks of relativistic accelerations on CP forces, we derive a general formula which allows for the computation of Casimir-Polder forces in generic stationary conditions from first principles, extending to fourth order in perturbation theory a method developed by Dalibard \emph{et al.} \cite{DDC}. In particular, we consider the interaction energy, arising from quantum vacuum fluctuations, among two atoms moving with a uniform proper acceleration $a$ in the same direction and separated by a constant distance $z$, perpendicular to their trajectories, and linearly coupled to a scalar field.
We show that the Casimir-Polder force between the two accelerating atoms displays a novel transition in its distance dependence at a new length scale, $z_a$, given by the inverse of the atomic acceleration (hereafter we use natural units $\hbar = c = k_B = 1$.). Such a transition is a cross-over in the interaction energy, from a Casimir-Polder potential for $az \ll 1$ , where the static zero-temperature interaction (as $z^{-2}$ and $z^{-3}$ in the nonretarded and retarded regimes, respectively) receives  a small thermal-like correction due to acceleration at the Unruh temperature $T_U=a/2\pi$, to a non-thermal interaction energy for $az \gg 1$, characterized by a $z^{-4}$ power law decay. This result should be compared with the Casimir-Polder force between two static atoms interacting with the scalar field at temperature $T$,
where at the thermal wavelength $\lambda^{th.}\sim1/T$  the interaction shows a transition from the $z^{-3}$ quantum regime to the $z^{-2}$ thermal classical regime.
The new characteristic length $z_{a} \sim 1/a$ is associated to the breakdown of the approximate description of the system in terms of a local inertial frame, and it indicates that the Casimir-Polder interaction is strongly reshaped by the presence of the non-inertial space-time background, associated to the relativistic accelerated motion of the two atoms.
This phenomenology is a simple non-trivial extension of the Unruh thermal response detected by a single accelerated observer to a system of two accelerated particles.

\emph{Casimir-Polder interactions.} We consider the Hamiltonian of a pair of two-level atoms ($A$,$B$), characterized by the same transition frequency $\omega_0$ and linearly coupled to a massless scalar field $\phi(x)$ by the coupling constant $\lambda$. The Hamiltonian can be written in the Dicke notation \cite{Dicke} and in natural units ($\hbar =c=1$), as
\begin{eqnarray}\label{Ham}
H&=&\omega_0\sigma^A_3(\tau)+\omega_0\sigma^B_3(\tau)+\int d^3k \omega_\textbf{k} a_\textbf{k}^\dag a_\textbf{k} \frac{dt}{d\tau} \nonumber \\
&+& \lambda\sigma^A_2(\tau)\phi(x^A(\tau))+\lambda\sigma^B_2(\tau)\phi(x^B(\tau)),
\end{eqnarray}
where $\sigma_i \, \,(i=1,2,3)$ are the Pauli matrices, $a_\textbf{k}$, $a_\textbf{k}^\dag$ are the annihilation and creation operators of the massless scalar field $\phi(x)$ with the linear dispersion relation $\omega_\textbf{k}=|\textbf{k}|$. The Hamiltonian \eqref{Ham} is expressed in terms of the same proper time $\tau$ of the two atoms (assuming a background flat spacetime), and the interaction term is evaluated on a generic stationary trajectory $x(\tau)$ of the two atoms. The distance $z$ between the atoms, perpendicular to their acceleration, is constant. Quantum fluctuations of the field, as well as radiation source fields, can induce an effective interaction among the two atoms at fourth order in the atom-field interaction.  Following the procedure proposed in \cite{DDC}, we can split the rate of variation $\frac{d\mathcal{O}^A}{d\tau}$ of an atomic observable $\mathcal{O}^A$ in the sum of two contributions, \emph{vf} and \emph{rr},
\begin{eqnarray}\label{vfc}
\left( \frac {d\mathcal{O}^A}{d\tau} \right)_{vf/rr}&=&\frac{i\lambda}{2}\Big(\phi^{f/s}(x_A(\tau))[\sigma_2(\tau),\mathcal{O}^A]+ \nonumber \\
&+&[\sigma_2(\tau),\mathcal{O}^A]\phi^{f/s}(x_A(\tau))\Big) \, ,
\end{eqnarray}
where the free term $\phi^f$ is the contribution present even in the absence of interaction, while the source term $\phi^s$ is the part due to the atom-field coupling and containing the coupling constant $\lambda$. The first contribution, $\left(d\mathcal{O}^A/d\tau\right)_{vf}$ describes the change in $\mathcal{O}^A$ caused by the fluctuations of the field that are present even in the vacuum (\emph{vacuum fluctuations}), while the second term represents the influence of the atom on the field, which in turn can act back on the atom (\emph{radiation reaction}). The method consists in rewriting  these contributions at a given order in perturbation theory  as quantum evolutions given by two effective Hamiltonians, $H_{eff}^{vf/rr}$, and then compute the \emph{vf} and ${rr}$ contributions to the atomic energy level shift (a second-order calculation is sufficient for the Lamb shift). In order to derive the Casimir-Polder interaction for two atoms ($A$,$B$), in the quantum states $|\alpha\rangle$ and $|\beta\rangle$ respectively, and moving with two arbitrary stationary trajectories $x_A(\tau)$ and $x_B(\tau)$ (the trajectories of the two atoms differ by a space translation only), we derive the effective Hamiltonians $H_{eff}^{vf/rr}$ at fourth order in $\lambda$ for one of the two atoms, as we shall report in detail elsewhere \cite{MNP}. We disregard the energy shifts independent from the atomic separation, because they do not contribute to the interatomic force.  We find the following expression of the \emph{vacuum fluctuations} contribution to the energy level shift of atom $A$ in the state $\mid \alpha \rangle$
\begin{widetext}
\be\label{vf}\begin{split}
\delta E_{\alpha, vf}^A=4i\lambda^4\lim_{(\tau-\tau_0)\rightarrow+\infty}\int_{\tau_0}^\tau d\tau'\int_{\tau_0}^{\tau'} d\tau''\int_{\tau_0}^{\tau''} d\tau'''&C^F(\phi^f(x_A(\tau)),\phi^f(x_B(\tau''')))\chi^F(\phi^f(x_A(\tau')),\phi^f(x_B(\tau''))) \\
&\times\chi_\alpha^A(\tau,\tau')\chi_\beta^B(\tau'',\tau'''),
\end{split}\ee
where we have introduced respectively the field symmetric correlation function and the field susceptibility,
\be\label{corr1}\begin{split}
C^F&(\phi^f(x_A(\tau)),\phi^f(x_B(\tau')))=\frac{1}{2}\langle 0|\{ \phi^f(x_A(\tau)),\phi^f(x_B(\tau'))\}|0\rangle,\\
\chi^F&(\phi^f(x_A(\tau')),\phi^f(x_B(\tau')))=\frac{1}{2}\langle 0|[\phi^f(x_A(\tau)),\phi^f(x_B(\tau'))]|0\rangle.
\end{split}\ee
\end{widetext}
Analogously, we have defined the atomic susceptibility of atom $A$ $(B)$ in state $\alpha$ $(\beta )$,
\be\label{corr2}
\chi_{\alpha/\beta}^{A/B}(\tau,\tau')\equiv\frac{1}{2}\langle \alpha/\beta|[\sigma_{2, A/B}^f(\tau), \sigma_{2, A/B}^f(\tau')]|\alpha/\beta\rangle.
\ee

In Eqs. (\ref{vf}-\ref{corr2}) the superscript $f$ stands for the free evolution of the operators. Although in \eqref{corr1} we have assumed the field in its ground state $|0\rangle$, Eq. \eqref{vf} is valid also for a quantum field at finite temperature, using the appropriate statistical functions; also, in the following we shall assume the two atoms prepared in their ground state $|g\rangle$.

The result \eqref{vf} has a sharp physical interpretation. The vacuum fluctuations contribution to the interatomic energy originates from
a non-local field correlation (expressed by $C^F$), present even in the vacuum state, which induces and correlates dipole moments in the two atoms ($\chi^{A/B}$), that eventually polarizes the field ($\chi^F$).
This physical picture is also consistent with the paradigmatic interpretation of Casimir-Polder forces by Power and Thirunamachandran in \cite{Power}, and often used to calculate also many-body Casimir-Polder forces \cite{CP97,Salam06,RPP07}. The radiation reaction contribution can be obtained similarly \cite{MNP}, but for brevity we do not report here its explicit expression. It describes the complementary physical mechanism, in which the atom $A$ has a fluctuating dipole moment ($C^A$) and it polarizes the field ($\chi^F$); a dipole moment is thus induced in the second atom ($\chi^B$) and it polarizes the field ($\chi^F$), which eventually acts back on atom $A$.
It is possible to show that the radiation reaction contribution is negligible compared to the vacuum fluctuation contribution for all the cases considered in this Letter, specifically, at small temperatures, i.e. for $T\ll\omega_0$, or, in the case of two uniformly accelerating atoms, for $a\ll\omega_0$ \cite{MNP}. Thus we concentrate on the vacuum fluctuations contribution only.

As a non-trivial test for our results, we can first calculate the scalar Casimir-Polder interaction energy $E_{CP}$ between two stationary atoms. Similarly to the electromagnetic Casimir-Polder case \cite{CasimirPolder}, we find a transition from a \emph{near zone} regime, defined by $\omega_0 z\ll1$, where $E_{CP} \simeq -\frac{1}{1024\pi^2}\frac{\lambda^4}{\omega_0}\frac{1}{z^2}$, to a \emph{far zone} regime, $E_{CP} \simeq -\frac{1}{512\pi^3}\frac{\lambda^4}{\omega_0^2}\frac{1}{z^3}$, defined for distances $\omega_0 z\gg1$, where retardation effects in the propagation of the field are relevant. A generalization of Eq. \eqref{vf} allows to obtain also the scalar Casimir-Polder force at finite temperature $T$, in terms of the thermal correlation function and susceptibility for a scalar field
\begin{widetext}
\begin{equation}\label{termico}\begin{split}
C^F_{th.}(\phi^f(x_A(\tau)),\phi^f(x_B(\tau')))&=\frac{1}{8\pi^2}\frac{1}{z}\int_0^\infty d\omega\sin(\omega z)\coth\Big(\frac{\omega}{2T}\Big) (e^{-i\omega(\tau-\tau')}+e^{i\omega(\tau-\tau')}),\\
\chi^F_{th.}(\phi^f(x_A(\tau)),\phi^f(x_B(\tau')))&=\frac{1}{8\pi^2}\frac{1}{z}\int_0^\infty d\omega\sin(\omega z) (e^{-i\omega(\tau-\tau') }-e^{i\omega(\tau-\tau')}).
\end{split}\end{equation}
\end{widetext}
The explicit computation is performed in the limit of small temperatures, $T\ll\omega_0$, following a general method originally introduced by Lifshitz \cite{Termico,Termico1,Termico2}. In view of the comparison with the Casimir-Polder force between two accelerated atoms, which is the main point of this Letter, it is important to stress that at finite temperatures, the massless thermal wavelength $\lambda^{th.}\sim1/T$ separates a quantum regime from a classical thermal regime. Indeed, for distances $z\ll\lambda^{th.}$ we find the expression for the static scalar Casimir-Polder force in near and far zone plus subleading thermal corrections proportional to
$ -\frac{\lambda^4}{z}(\frac{T}{\omega_0})^2$; on the other hand, for distances larger than the typical length scales associated to quantum effects, i.e. for $z\gg\lambda^{th.}$, the Casimir-Polder force manifests again a classical thermal behavior similar to that in the near zone
\be\label{thermalCP}
E_{CP}^{ th.}=-\frac{1}{512\pi^3}\frac{\lambda^4}{\omega_0^2}\frac{T}{z^2},
\ee
as it has been already noticed for the electromagnetic case \cite{Barton01,Termico2}.

\emph{Unruh corrections to Casimir-Polder interactions.} We now apply our result \eqref{vf} to the case of two atoms moving with the same uniform acceleration, perpendicular to their separation. In this case, a modification of their Casimir-Polder interaction is expected, because the two atoms perceive modified vacuum fluctuations, as the Unruh effect would suggest. We have already obtained hints of such modification \cite{MP10} and discussed observability of this new phenomenon, that could be a way to confirm experimentally the evidence of the Unruh effect \cite{NP13}. An atom moving with uniform relativistic acceleration $a$ in the $\hat{x}$ direction follows the worldline
\be\label{line}
t(\tau)=\frac{1}{a}\sinh(a\tau) \quad x(\tau)=\frac{1}{a}\cosh(a\tau) \quad y(\tau)=z(\tau)=0.
\ee

We are now going to show how interatomic Casimir-Polder interactions allow to distinguish the effect of a relativistic acceleration from a thermal behavior. Even if such a thermal character have been envisaged in a large number of situations \cite{CHM08,SCD81,AM94,HZ07}, departures from thermal predictions for accelerating atoms have been shown in the Lamb shift and in the spontaneous excitation of accelerating atoms, coupled to the electromagnetic field, in vacuum space \cite{Passante98, Takagi86} or in front of a conducting plate \cite{Rizzuto07,YL05,ZYL06,RS09}.

In such a situation it is convenient to introduce a new set of coordinates, necessary to cover the Minkowski spacetime $(t,x)$ accessible to accelerated observers. They are defined in two regions, the Rindler wedges, which are causally disconnected, and where a Rindler metric can be defined accordingly \cite{CHM08, BD}.

We consider two uniformly accelerating atoms, moving along the worldlines \eqref{line} with the same uniform acceleration $a\ll\omega_0$, and separated by a distance $z$ orthogonal to the acceleration direction $\hat{x}$.
We show that, at short distances, Casimir-Polder interactions can probe thermal Unruh-like effects, while at larger distances they reveal a non-thermal behavior due to the intrinsically non-inertial nature of the Rindler metric. As done in \eqref{termico} for the thermal Casimir-Polder force, we first obtain the correlation function and susceptibility of the scalar field in the accelerated background
\begin{widetext}
\begin{equation}\label{corrRindler}\begin{split}
C^F_{acc.}(\phi^f(x_A(\tau)),\phi^f(x_B(\tau')))&=\frac{1}{8\pi^2}\frac{1}{\mathcal{N}(z,a)}\int_0^\infty d\omega f(\omega,z,a)\coth\Big(\frac{\pi\omega}{a}\Big) (e^{-i\omega(\tau-\tau')}+e^{i\omega(\tau-\tau')}) \, ,\\
\chi^F_{acc.}(\phi^f(x_A(\tau)),\phi^f(x_B(\tau')))&=\frac{1}{8\pi^2}\frac{1}{\mathcal{N}(z,a)}\int_0^\infty d\omega f(\omega,z,a) (e^{-i\omega(\tau-\tau') }-e^{i\omega(\tau-\tau')}) \, ,
\end{split}\end{equation}
\end{widetext}
where $f(\omega,z,a)=\sin(\frac{2\omega}{a}\sinh^{-1}(\frac{az}{2}))$ and $\mathcal{N}(z,a)=z\sqrt{1+(az/2)^2}$. A close comparison between \eqref{corrRindler} and \eqref{termico} shows that for $az\ll1$ the correlation function \eqref{corrRindler} has a thermal-like behavior set by the Unruh temperature, $T_U$. Hence, the vacuum fluctuations contribution \eqref{vf} to the Casimir-Polder interaction exhibits, at the lowest order in $az$,
the same thermal-like correction $\sim -\frac{\lambda^4}{z}(\frac{T_U}{\omega_0})^2$,
found for the Casimir-Polder interaction at finite temperature. At higher orders in $az$, Eq. \eqref{corrRindler} shows that the correction due to the accelerated atomic motion starts to differ significantly from the correction due to a finite temperature \cite{NotaCorr}. This discrepancy suggests a strong breakdown of the usual analogy between acceleration and finite temperature effects for the Casimir-Polder potential at distances $z\gg z_a \sim 1/a$, resulting in a novel power law behavior of the Casimir-Polder interaction,
\be\label{acceleratoCP}
E_{CP}^{acc.}=-\frac{1}{512\pi^4}\frac{\lambda^4}{\omega_0^2}\frac{z_a}{z^4}.
\ee

\begin{figure}[h]\centering
\includegraphics[width=8.5cm]
{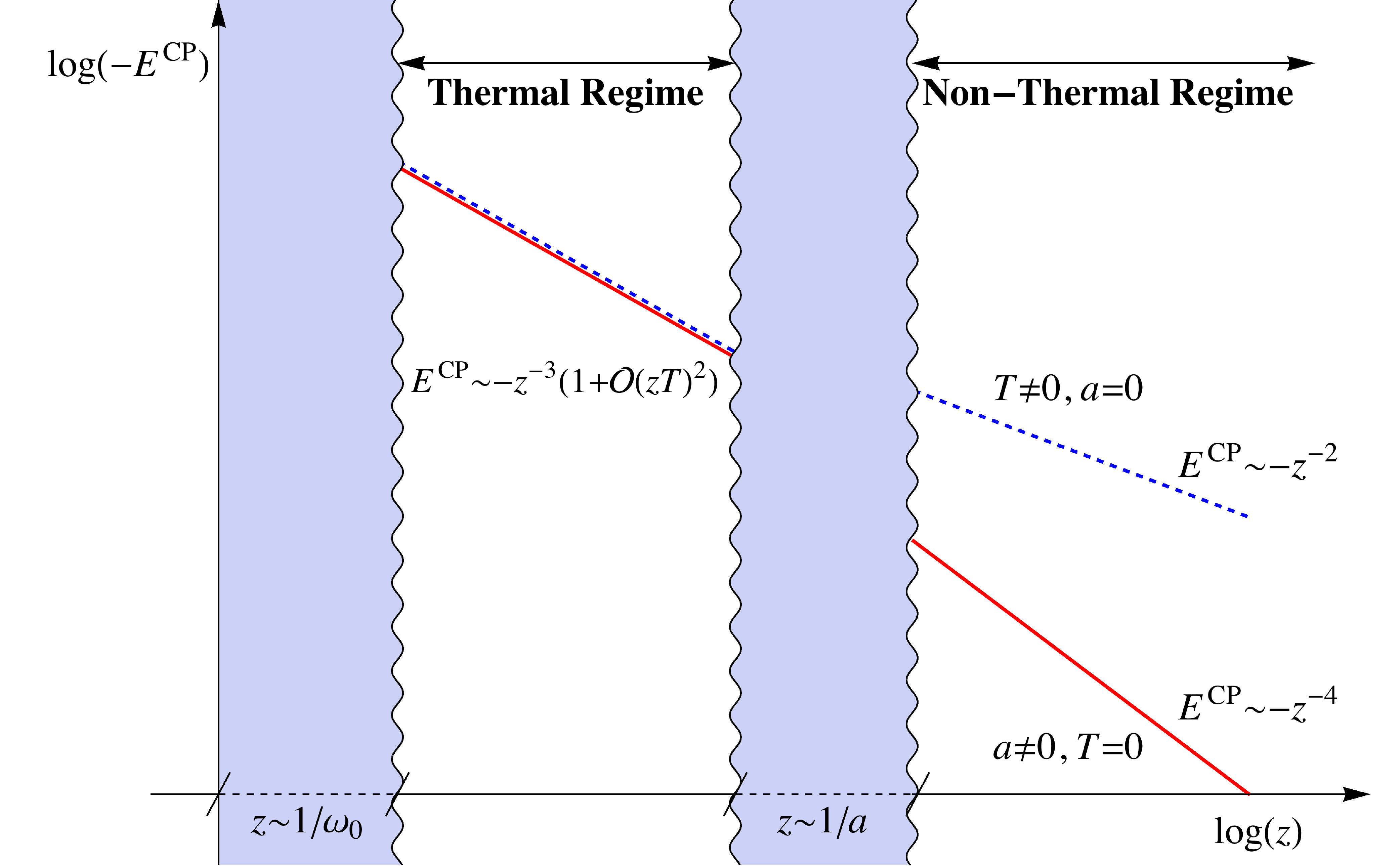}
\caption{(Color online) Comparison between the Casimir-Polder interaction among two atoms moving with relativistic uniform acceleration $a$ and constant separation $z$ (red continuous line), and the static interaction for atoms at rest at temperature $T=a/2\pi$ and same distance (blue dashed line), in far zone, $z \gg 1/\omega_0$. While for short distances, $z \ll 1/a$, both potentials display the same thermal-like behavior, at distances larger than the characteristic length scale $1/a$, the thermal and the accelerated Casimir-Polder potentials exhibit a sharply different power law decay with the interatomic distance.}
\label{Fig1}
\end{figure}

Our result \eqref{acceleratoCP} shows that the Casimir-Polder interaction energy between two accelerated atoms decreases faster with the distance than in both near and far zones. This can be guessed from the following heuristic argument: since both atoms are accelerating, the distance traveled by a scalar photon emitted by one atom to reach the other atom increases with time, and this results in an overall decrease of the interaction strength among them \cite{NP13} (a more precise comparison between our result in \eqref{acceleratoCP} and the results in \cite{NP13} is not straightforward because in the latter case the interaction energy is time-dependent and valid in a well-defined time interval, while the present result involves a time average of the interaction energy, as it is evident from \eqref{vf}). We can consider the behavior described by Eq. \eqref{acceleratoCP} as a new quantum regime, as opposed to the classical thermal regime given by Eq. \eqref{thermalCP}, which on the contrary destroys the quantum retarded Casimir interaction decaying as $z^{-3}$.
Also, we wish to stress that the distance $z_a$ is the characteristic length scale for the breakdown of the local inertial frame approximation \cite{MTW73}: for distances smaller than $z_a$, it is possible to find a local inertial frame where the correlation functions of the scalar field are fairly well described by the their thermal Minkowski analogue, and the only net effect of acceleration is embodied in the Unruh thermal analogy; on the other hand, signals spreading over distances larger than $z_a$ must take into account the non-inertial character of relativistic acceleration, encoded in the non-Minkowskian metric.
Consequently, field quantization in Rindler spacetime will strongly affect the nature of vacuum fluctuations ($C_F$) and field susceptibility ($\chi_F$), ultimately leading to the novel power law behavior of the Casimir-Polder potential \eqref{acceleratoCP}. This phenomenology is in sharp contrast with the \emph{classical} effect outlined above for the Casimir-Polder interaction at finite temperature (see Eq. \eqref{thermalCP}) and it is summarized in Fig. \ref{Fig1}. It should be noted that such an effect cannot be detected by a single uniformly accelerated \emph{point-like} detector in the unbounded space, as in \cite{CHM08,AM94}, since in that case it is always possible to find a local set of Minkowski coordinates in the neighborhood of a point-like detector. With this respect, our result can be seen as a simple non-trivial extension of the Unruh thermal response detected by a single accelerating observer, to a system of two relativistic accelerated systems.
Finally, we would like to point out that the qualitative change of Casimir-Polder force described by Eq. \eqref{acceleratoCP} is ultimately grounded on the non-inertial character of the accelerated background and it is expected to manifest ubiquitously also for other fields, as well as for a multi-level atom configuration.

\emph{Conclusions and Perspectives.} In this Letter we have shown how Casimir-Polder forces among two uniformly accelerating atoms can probe non-thermal effects beyond the Unruh analogy between uniform acceleration and finite temperature.
We have shown that for interatomic distances above the characteristic length scale associated to a local inertial description of the system, the Casimir-Polder energy shows a different power law dependence with the distance, compared to the corresponding potential at finite temperature.

A qualitative change of the interatomic potential may affect some macroscopic properties of an accelerated many-atoms system, as the following example would suggest (analogous ideal experiments were envisaged in \cite{UW82} for an accelerating box filled with photons). Let us consider a box filled with atoms with a given proper density and moving with finite acceleration $a$. The qualitative change of the interaction between the atoms from a marginal long-range $z^{-3}$ to a short-range $z^{-4}$ given by \eqref{acceleratoCP} at the acceleration-dependent scale $z_a$, could manifest in a change of the thermodynamics properties (for example in the equation of state of the gas), if the average interatomic distance is larger than $\sim 1/a$, since thermodynamics of long-range and short-range interacting systems is sharply different \cite{Ruffo2009}. This density/acceleration cross-over is of quantum origin, and it could have also consequences on thermodynamics of the Universe during the stages of its evolution.

Also, our new expressions for the fourth-order vacuum fluctuations and radiation reaction contributions to energy shifts have a general validity, and they could be straightforwardly applied to investigate electromagnetic dispersion interactions involving accelerating atoms \cite{Boyer84,Takagi86,Passante98,RS09, NP13} or atoms in circular motion, which could be relevant to detect the Unruh effect \cite{BL83}. Furthermore, they can be easily employed to compute  dispersion forces among two atoms outside a Schwarzschild black hole or in de Sitter spacetime, where Casimir forces could provide new physical insights into problems of cosmological interest (similarly to recent computations of Lamb Shifts in  curved backgrounds \cite{Schwarz}).

\emph{Acknowledgments.} The authors wish to thank A. Gambassi, T. Langen, B. Rauer, L. Rizzuto and S. Spagnolo for stimulating discussions on the subject of this paper, and I. Carusotto and S. Liberati for a critical reading of the manuscript and useful suggestions. We wish also to thank the anonymous Referee for his/her valuable suggestion on thermodynamics, discussed in the conclusions. The authors gratefully acknowledge financial support by the Julian Schwinger Foundation, by MIUR and CRRNSM.


\begin{thebibliography}{120}
\bibitem{NotaCorr} A similar behaviour of the correlation functions \eqref{corrRindler} can be extrapolated from the Lamb shift and spontaneous emission corrections of an accelerated atom near a conducting plate obtained in Refs. \cite{Rizzuto07,YL05,RS09}.
\end{thebibliography}
\end{document}